\documentclass[12pt,prd]{revtex4}
\usepackage{dsfont,amssymb,amsmath}
\usepackage{graphicx}
\usepackage{epstopdf}
\usepackage[mathscr]{euscript}

\newcounter{aaa}

\newenvironment{teor}[2][{}]{\begin{trivlist}\refstepcounter{aaa}%
\labelsep=0pt\item[\bfseries \theaaa. #2.
]#1}
{\end{trivlist}} \newcommand*{\polu}{P} 
\newcommand{\rmd}{\mathrm{d}}
\newcommand{\obozn}{\equiv} 
\newcommand{\evalat}[3]{\left.#1\right|_{#2}^{#3}}

\newcommand{\ssy}[5]{#1,    #2 {\bf #3}, #5 (#4)\rlap{.}}

\newcommand*{\Hi}{\ensuremath{\EuScript{H}}}
\newcommand*{\Ci}{\ensuremath{\EuScript{C}}}
\newcommand*{\Cio}{\ensuremath{\mathring{\EuScript{C}}}}
\newcommand*{\Fi}{\ensuremath{\EuScript{F}}}

\newcommand*{\KG}[1]{(#1)_\text{KG }}
\newcommand*{\Sc}[1]{<#1>}

\begin{document}
\title{Finite energy  quantization  on a topology changing spacetime}
\author{S. Krasnikov\thanks{Email: krasnikov.xxi@gmail.com}}%
\affiliation{Central Astronomical Observatory at Pulkovo, St.Petersburg, 196140, Russia}
\date{}
\begin{abstract}
    The ``trousers'' spacetime is a  pair of flat 2D cylinders (``legs'') merging into  a single one (``trunk''). In spite of its simplicity  this     spacetime has a few features (including, in particular, a naked singularity in the ``crotch")
each of which is presumably  unphysical, but for none of which a mechanism is known able to prevent its occurrence.
 Therefore it is interesting and important to study the behavior of the quantum fields  in such a space.   Anderson and
 DeWitt were the first to consider the free scalar field in the trousers spacetime. They argued   that  the crotch singularity produces an infinitely bright
flash, which was interpreted as evidence that the topology of space is dynamically preserved.  Similar divergencies were
later discovered by Manogue, Copeland and Dray who used a more exotic quantization scheme. Later yet the same result
obtained within a somewhat different approach led  Sorkin to the conclusion that the topological transition in question is
suppressed in quantum gravity.

In this paper I show that the  Anderson--DeWitt divergence is an artifact of their choice of the Fock space. By
choosing  a different one-particle Hilbert space one gets a quantum state in which the  components of the stress-energy tensor (SET) are bounded in the frame of a free-falling observer.

\end{abstract}
\maketitle

\section{Introduction and conclusions} 
\begin{figure}
\includegraphics[width=\textwidth]{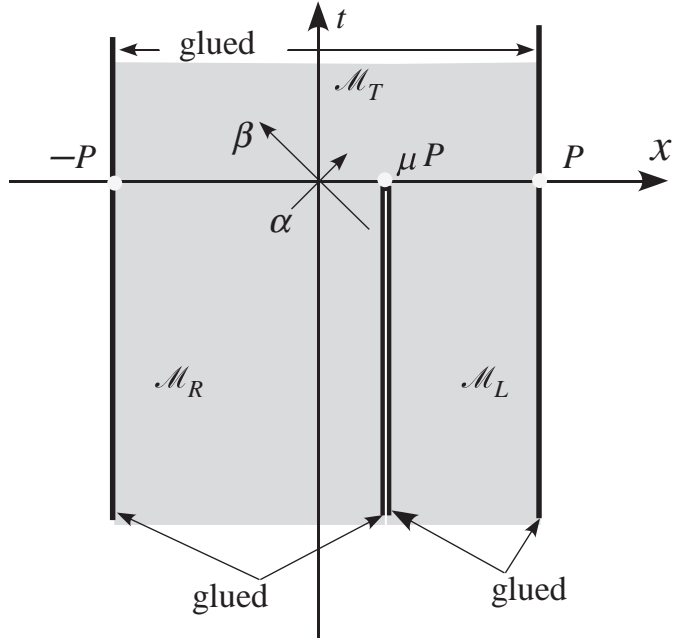}
  \caption{Constructing the trousers spacetime from a flat strip. The
  left leg
$\mathcal{M}_L\obozn \Bigl(t\leq 0,\   x\in [-\polu, \mu\polu)\Bigr)$, the right leg $ \mathcal{M}_R\obozn \Bigl(t\leq0,\  x\in [\mu\polu ,\polu)\Bigr)$, and the trunk $\mathcal{M}_T\obozn \Bigl(t\geq 0,\  x\in (-\polu ,\polu]\Bigr)$ are obtained by gluing together the rays bounding the  corresponding strips.
  The white circles depict the removed points. They cannot be returned back after the relevant identifications are performed and thus a naked quasiregular singularity appears.
\label{fig:space_1}}
\end{figure}

The two-dimensional spacetime $\mathcal M$ called ``trousers"  is  obtained from the strip \[ \rmd
s^2 = \rmd t^2  - \rmd x^2\qquad t\in \mathds{R},\quad x\in [-\polu,\polu] \] by, first, deleting the points $t=0,\ x=\pm
\polu$ and  the ray $t\leq 0,\ x=\mu \polu$, then attaching a copy of the deleted ray to either bank of the cut and, finally, smoothly
gluing each of the six rays to its counterpart
so that the resulting space consists of three cylinders, see figure~\ref{fig:space_1}.

The trousers spacetime merits the most detailed consideration because in spite of its simplicity it possesses two features,
interesting and important, but poorly understood:
\begin{enumerate}
  \item the topology of its spacelike sections  changes with time. It is $\mathds{S}^1 \sqcup\mathds{S}^1$ at negative $t$ and
      $\mathds{S}^1$ at positive. This type of topology change is particularly significant, because it
      may have to do with the  appearance of a wormhole or (if the $t$-axis is directed to the past as in \cite{AdW}) with the final stage of the wormholes (including the Schwarzschild black hole) evaporation
      \cite{stro};
  \item the spacetime is singular, as one might expect, and the singularity---loosely speaking it is located at the ``crotch of
      the trousers"--- is naked and quasiregular. Presumably either of these  properties  makes  it ``unphysical", but no
      mechanism is found that would protect the Universe from the appearance of such singularities.
\end{enumerate}

The evolution of a quantum field in trousers was first considered by  Anderson and  DeWitt (AD). In their well-known paper
\cite{AdW} they conjectured  that the above-mentioned singularity emits an infinitely bright flash. Their reasoning was as
follows \footnote{With \emph{our} choice of the future direction the mode's labels `in' and `out' must be interchanged.} ``[\dots] an `in' mode function propagating to the right splits into components propagating to the right in each
leg. Although continuous in the trunk region, such mode functions generally have discontinuities [\dots] in the legs. [\dots]
Every `out'  mode function is continuous in each leg (vanishing in one of them) but has discontinuities in the trunk region.
[\dots] When these functions are differentiated the discontinuities give rise to delta functions. Since the terms of the mode
sum for $<\text{in, vac}| \mathbf{T}_\text{ren}^{00} |\text{in, vac}>$ are bilinear in differentiated mode functions,
the square of the delta function automatically appears''.

The divergence of $<\text{in, vac}| \mathbf{T}^\text{ren}_{00} |\text{in, vac}>$ does not automatically exclude the
topology  changes: it may happen that \emph{some}  of them are free from that divergence \cite{DowSur} or that the
relevant quantity is the matrix element $<\text{in, vac}| \mathbf{T}^\text{ren}_{00} |\text{out, vac}>$ as opposed to
the expectation value of the stress-energy tensor \cite{forks}. It is also possible that the fields in such an unusual space
must be quantized in some special way  \cite{barv} (one such  unusual quantization was proposed in  \cite{MCD}, the
resulting $< \mathbf{T}^\text{ren}_{00}>$, though, diverges all the same). There is a good consensus, however, that
the AD flashes are an indicator of some ``flaw" in the trousers spacetime  \cite{AdW}, \cite{IshHo}, \cite{Duston}.
The goal of this paper is to show that this is not the case.
\subsection*{Conclusions}
In regard to the divergence of the energy density the trousers
turn out to be  as ``nice" as, say, the Schwarzschild space. The latter is not compromised by the fact that in \emph{some} states
(such as the Boulware vacuum) the energy density diverges at the horizon. What matters is the existence of states
\emph{free} of
 such divergences. Accordingly, we rehabilitate the trousers spacetime by explicitly constructing a state in which
$< \mathbf{T}^\text{ren}_{ik}>$ are bounded. Note that the existence of such a state does not contradict to the
argument quoted above, owing  to the word ``generally" used in the latter.

\section{The quantization}

\subsection{The plan}\label{sub:plan}
The field $\phi$ considered in this paper  obeys (classically)  the wave equation \begin{equation}\label{eq:wave}
    \Box \phi=0.
\end{equation}
Though the spacetime under discussion is non-globally hyperbolic, the piecewise smooth (see below) complex-valued solutions of
\eqref{eq:wave}
 are fixed uniquely---this is proven in section~\ref{sec:reduc}---by  the data at any surface
$t=const\leq 0$. In this sense  the singularity is ``harmless" \cite{IshHo,CauNGH} and we can (and shall) proceed exactly as in the
globally hyperbolic case. In doing so we are guided by  the textbook \cite{BirDav}, in particular, the
units and sign conventions are those used there.

To canonically quantize the field one  must first expand it as a series in  vacuum modes  that is
find a set of functions  $\{\phi_k\}$ on $\mathcal{M}$ that are   an orthonormal basis in a  Hilbert space ${\Hi}$. It is the choice of ${\Hi}$ that encodes the physics of the problem and determines the resulting theory.

The field operator $\hat \phi$ in QFT is taken to be an (operator-valued) \emph{distribution}. But the Hilbert space in discussion is usually built on the basis of \emph{smooth} solutions of the classical equation of motion (that is, \Hi\ may contain non-smooth functions, but only those to which a sequence of smooth ones converges). Such a choice seems inadequate in studying ``thunderbolts" with their discontinuities. Therefore Manogue, Copeland and Dray in \cite{MCD} expanded the space by allowing the vacuum modes to have jumps. However, the derivatives  of such modes will have $\delta$-like singularities exactly where the modes are discontinuous, which makes the Klein--Gordon  scalar product  ill-defined, see \eqref{eq:scalPr}. Physically  such singularities seem unwarranted too. Indeed, the crotch singularity can play the role of a source, so the solutions to the (now inhomogeneous) wave equation are expected to have irregularities on the null geodesics emanating from the ``missing point". However,   the energy density   proportional to the square of the delta function is a \emph{too} strong irregularity. So, in this paper we propose an intermediate approach and require classical solutions to be \emph{continuous}, though not continuously differentiable.
Specifically, let $\Ci$ be the space of  bounded continuous
complex-valued
functions on $\mathcal{M}$
that are smooth   solutions to  equation~\eqref{eq:wave} on the whole $X$ except perhaps at the points of  past incomplete inextendible  null geodesics (i.~e., loosely speaking, null geodesics emanating from the singularity), where the  derivatives of those functions may have jump discontinuities.
Correspondingly,   we are looking for a space    \Hi\ such that
\begin{enumerate}
 \item \Hi\ is a Hilbert space with respect to the Klein--Gordon   scalar product   \begin{equation}\label{eq:scalPr}
    \KG{f,g}\equiv i\int_{-\polu}^{\polu}\evalat{(g^*\dot f -f{\dot g}^*)}{t=0}{}\,\rmd x
\end{equation} (we indicate the particular value of $t$ because at this stage we cannot guarantee that the integral does
not depend on the choice of that value, the spacetime being non-globally hyperbolic). The positive-definiteness of this form
is a non-trivial restriction on  \Hi.
\item  up to a constant any function $f$ in  $\Ci$ is the sum of a function from \Hi\
and a function from~$\Hi^*$:
 \begin{equation}\label{eq:poln} \text{for all } f \in \Ci
\qquad f=f^++(f^-)^* + c,\qquad f^+,f^-\in \Hi,\quad c=const. \end{equation} $f^+ $ and $(f^-)^*$ are often called, respectively,
``positive and negative  frequency'' parts of~$f$. Conversely, \Hi\ must not include ``superfluous functions",  in other words,  \Hi\ must not have a proper subspace  satisfying \eqref{eq:poln}.
\end{enumerate}
\begin{teor}{Remark}\label{eq:arrow}
    The word \emph{past} in the definition of \Ci\ signifies some time asymmetry in our approach  which is not related to the asymmetry of the underlying spacetime.
\end{teor}
\subsection{ Reduction to initial conditions}\label{sec:reduc} In this subsection we represent in a    convenient form the space \Ci: we  use the periodicity of the functions constituting  $\Ci$ to express them in terms of  their restrictions to  the surface $t=0$ (thus trading functions of two variables each for a pair of functions of one variable).

To begin with we note that any
$ \phi\in \Ci$  being a solution to the wave equation
\[
\partial_\alpha\partial_\beta \phi=0,\qquad \alpha\obozn  t+x,\quad  \beta \obozn t-x
\]
is the sum of a right-moving and a left-moving (i.~e. depending---within each of the cylinders---on the $\alpha$- or, respectively, $\beta$-coordinate of its argument) function. Put more formally, it has the form
\begin{equation}\label{eq:phi}
  \evalat{\phi}{\mathcal{M}_\curlywedge }{}(p)= a_\curlywedge (\alpha(p)) +   b_\curlywedge (\beta(p))+  c ,
\qquad\curlywedge\obozn L,R,T.
\end{equation}
Here $c$ is an arbitrary constant and
 $a_\curlywedge (\alpha)$, $b_\curlywedge (\beta)$ for each value of $\curlywedge$ are a pair of  functions such that, first,
\begin{subequations}
\begin{equation}\label{eq:restr}
a_\curlywedge (\alpha(p)) =  \evalat{a}{\mathcal{M}_\curlywedge }{}(p),  \quad
b_\curlywedge (\beta(p)) =  \evalat{b}{\mathcal{M}_\curlywedge }{}(p),\qquad\text{where }
a,b\in\Ci
\end{equation}
[note that the  \emph{entire} functions $a(p)$, $b(p)$ do not have to be of the form $a(\alpha(p))$ and $b(\beta(p))$; moreover, they may have a discontinuity on the ray $t\leq 0,\ x=\mu \polu$] and, second, they satisfy the following  normalization conditions
 \begin{equation}\label{eq:normA}  a_L(\mu\polu)= \frac{\int_{-\polu}^{\mu\polu}a_L(\alpha)\,\rmd\alpha}{(1+\mu)\polu}
+\frac{\int_{\mu\polu}^{\polu}a_R(\alpha)\,\rmd\alpha}{(1-\mu)\polu},
\qquad
  b_L(\mu\polu)=\frac{\int_{-\polu}^{\mu\polu}b_L(\beta)\,\rmd\beta}{(1+\mu)\polu}
+\frac{\int_{\mu\polu}^{\polu}b_R(\beta)\,\rmd\beta}{(1-\mu)\polu}.
\end{equation}
The reason for choosing these particular conditions will become clear later, see \eqref{eq:S'}; for now notice only that for any $\phi$ eqs.~(\ref{eq:phi}--\ref{eq:normA}) define a unique $c$. The subset of $\Ci$ consisting of all functions
 $ \phi$ for which $c=0$ is denoted $\Cio$.

 Finally, the topology of our spacetime requires $ \phi $ to have some periodicity properties. In order to satisfy them we take   $ a_\curlywedge $ and $b_\curlywedge $ to be   periodic    functions of $\alpha$ and $\beta$, respectively \footnote{The
 term proportional to $\alpha +\beta$ is duly smooth on $\mathcal{M}$ without being a sum of such periodic functions; it is excluded, however, by the boundedness of $ \phi $.}. The period
\begin{equation}\label{eq:h}
\text{of  $a_L $ and $b_L$ is } (1+\mu)\polu,\qquad \text{of $a_R$ and $b_R$  is } (1-\mu)\polu,\qquad
\text{of $a_T$ and $b_T$  is } 2\polu.
\end{equation}
\end{subequations}

Now let us introduce the aforementioned functions of one variable. To this end
denote by  $  \Fi$  the space of all continuous functions $A(x)\colon\ [-\polu ,\polu]\to \mathds{C}$ (it is convenient to imagine $A$ as defined on the surface $t=0$; in doing so one, strictly speaking, must keep in mind that this surface lacks the points $x=\pm \polu, \mu\polu$, we shall omit this trivial reservation from now on)
which
\begin{enumerate}
  \item are smooth, except, perhaps, at the points $x=\mu \polu$ where  the  derivatives are allowed to have  jump discontinuities;
 \item satisfy the condition
 \begin{equation}\label{eq:normF}
 A(\mu\polu)= \frac{1}{(1+\mu)\polu}\int_{-\polu}^{\mu\polu}A(x)\,\rmd x
+\frac{1}{(1-\mu)\polu}\int_{\mu\polu}^{\polu}A(x)\,\rmd x;
 \end{equation}
  \item obey the ``periodicity condition"
\begin{equation}\label{eqLperiodA}
A^{(n)}(-\polu) = A^{(n)}(\mu\polu-0),\quad A^{(n)}(\mu\polu+0)= A^{(n)}(\polu),\qquad n=0,1\ldots
\end{equation}
\end{enumerate}

Each pair $A,B\in   \Fi$ defines uniquely a function $\Psi(A, B)\in \Cio$ in the following way: 
$a_\curlywedge $ are defined to be the extensions by periodicity, see \eqref{eq:h},  of the functions, respectively,
\begin{equation*}
a_T\obozn A\quad \text{at }x\in (-\polu ,\polu], \qquad
a_L\obozn A\quad \text{at }x\in (-\polu, \mu\polu], \qquad
a_R\obozn A\quad \text{at }x\in  (\mu\polu ,\polu].
\end{equation*}
The functions $b_\curlywedge $ are dealt with in exactly the same manner [the only difference is in the sign: $\evalat{b_\curlywedge (\beta)}{t=0}{}=B(-x)$]. Now $\evalat{\phi}{\mathcal{M}_\curlywedge }{}$ are built by \eqref{eq:phi} with $c=0$ and, finally, $\Psi(A, B)$ is defined to be the result of gluing together all three restrictions $\evalat{\phi}{\mathcal{M}_\curlywedge }{}$.

Conversely, any $\phi\in \Cio$ defines uniquely a pair $A,B\in   \Fi$ such that $\Psi(A, B)=\phi$. This is done by decomposing
 $\phi$ into the right-moving and the left-moving parts $a(p)$ and $b(p)$, see \eqref{eq:restr}, and defining $A$, $B$ to be their restrictions
 \[
 A(x(p)) \obozn \evalat{a(p)}{t=0}{}, \qquad B(x(p)) \obozn \evalat{b(p)}{t=0}{}.
 \]
 We have thus established that $\Psi\colon \Fi\otimes \Fi\to \Cio $ is an isomorphism. It can be transformed into an isometry by an appropriate choice of the inner product in $\Fi$. Indeed, substituting the obvious  expressions
\begin{equation}\label{eq:data}
\evalat{\phi }{t=0}{}(x)= A(x)+ C(x),\qquad
\evalat{\dot\phi }{t=0}{}(x)= A'(x)- C'(x), \qquad\text{where } C(x)\obozn B(-x)
\end{equation}
into \eqref{eq:scalPr}
one finds
\begin{gather}\label{eq:scalProdP}
\KG{\phi_1,\phi_2}= \Sc{A_1,A_2} + \Sc{B_1,B_2} ,
\\\label{eq:scalProdSc}
\text{where}\quad \Sc{F_1,F_2}\obozn i\int_{-\polu}^{\polu} [F^*_2(x)F'_1(x) - {F'}^*_2(x)F_1(x)]\, \rmd x.
\end{gather}

\begin{teor}{Remark}\label{rem:harml}
    The formulas \eqref{eq:data} show that the singularity in $\mathcal M$ is harmless in the sense of \cite{IshHo}: there exists a unique solution to the wave equation for any Cauchy
data fixed at a surface $t=t_0\leq 0$. It is, of course, this property that enables one to study QFT on $\mathcal M$ without adopting a number of  \emph{ad hoc} assumptions.
\end{teor}

\subsection{The choice of the Hilbert space and its basis}

Let $\{u_{Lk}\}$ and $\{u_{Rk}\}$, $k=1,2,\ldots$ be the sets of functions defined  as follows:
\[u_{Lk}(x)\obozn \left\{
               \begin{array}{llll}
                 (4\pi  k)^{-\frac12}e^{-ik\xi}  & \hbox{at\ } x\in[-\polu,\mu\polu] & \text{ i. e. \ }
\xi  \in[-\pi, \pi];&\\
                 (4\pi  k)^{-\frac12}e^{-ik\pi}, & \hbox{at\ } x\in[\mu\polu,\polu],&\qquad\text{where }\xi  \obozn \tfrac{ \pi  }{1+\mu}(\tfrac{2}{\polu } x +1-\mu).&
               \end{array}
             \right.
\]
and likewise
\[u_{Rk}(x)\obozn \left\{
               \begin{array}{llll}
(4\pi  k)^{-\frac12}e^{-ik\zeta}  &\hbox{at\ } x\in[\mu\polu,\polu] & \text{ i. e. \ }
\zeta \in[-\pi, \pi],&
\\
                 (4\pi  k)^{-\frac12}e^{-ik\pi}, & \hbox{at\ } x\in[-\polu,\mu\polu],&\qquad\text{where }\zeta  \obozn \tfrac{ \pi  }{1-\mu}(\tfrac{2}{\polu } x -1-\mu).&\end{array}
             \right.
\]
The set $\{u_{Rk},u_{Lk}\},\ k=1,2,\ldots$ is
 orthonormal w.r.t. the scalar product $\Sc{,}$ defined in \eqref{eq:scalProdSc}
and we denote by $\Hi_F$ the (auxiliary) Hilbert space obtained by declaring that set to be a basis.

The  modes $\phi_k$ we are after are now defined as
\begin{equation}\label{eq:modes}
     \phi_{ Q k}\obozn  \left\{
                        \begin{array}{ll}
                           \Psi(u_{ Q k},0), & k> 0 \\
                           \Psi(0,u^*_{ Q |k|}), & k<0
                        \end{array}
                      \right.,\qquad \text{where }  Q\obozn R,L
\end{equation}
(i. e. $\phi_{ Q k}$ are obtained from $u_{ Q k}$ by replacing $x\dashrightarrow -\beta=x-t$ for $k< 0 $ and $x\dashrightarrow \alpha=x+t$ for $k> 0 $ and extending the resulting functions by periodicity). So, every   mode  is at first a harmonic wave that moves in the corresponding  leg (left or right depending on whether the subscript is $ L$  or $R$) to the left or to the right depending on the sign of $k$. In the trunk, however, the behavior of the mode becomes more exotic.  It is just a constant here except in a spiral strip bounded by two null geodesics emanating from the crotch. Within that strip the mode is still a piece of a harmonic wave whose fronts are just those geodesics. The mode, though continuous---which enables the SET to remain bounded, contrary to  the AD conjecture---is not smooth. So, one does not expect the energy to be conserved, but this is natural for  a non-static spacetime.
\begin{figure}
\includegraphics[width=\textwidth]{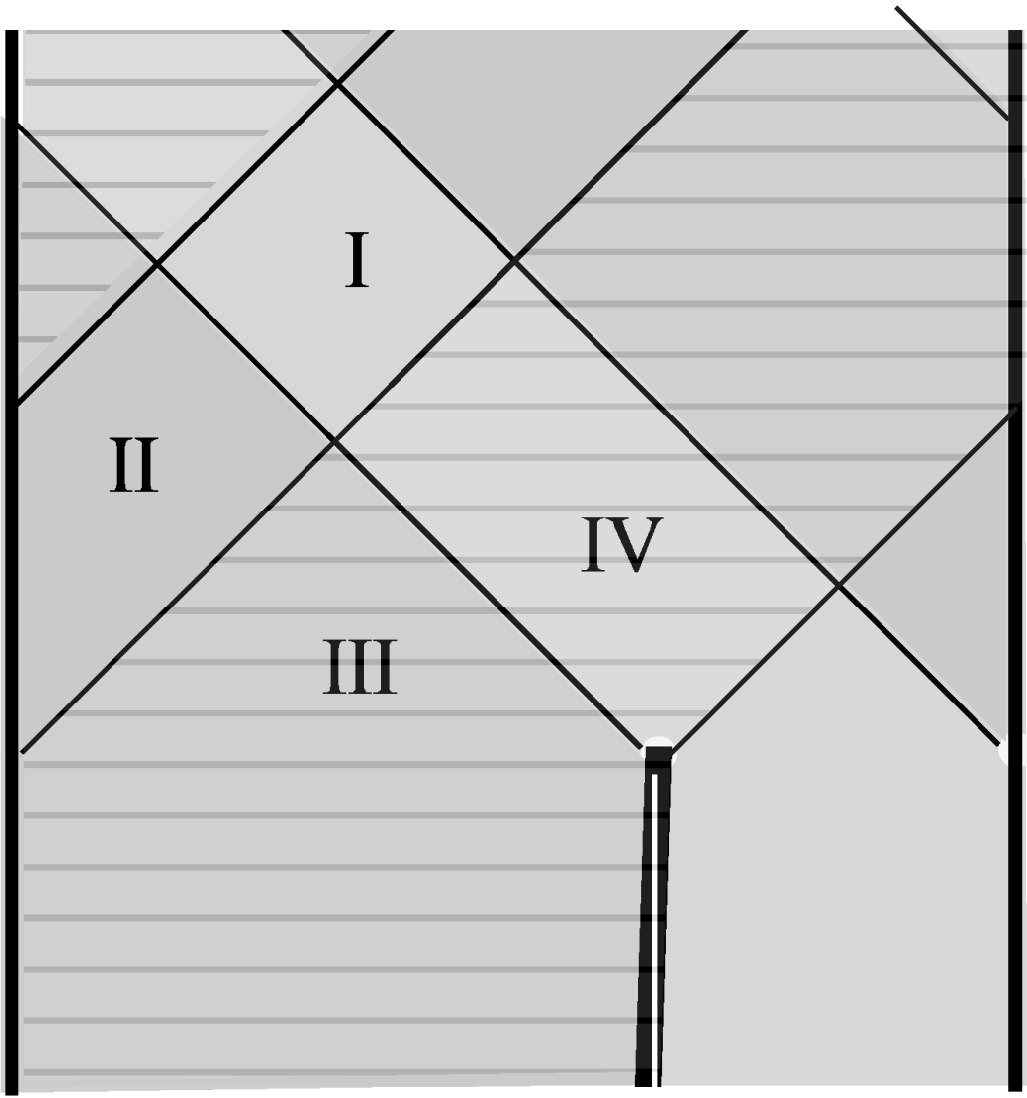}
  \caption{The slanted rays are the null geodesics at which the derivatives of functions of \Cio\ are allowed to have jumps. In the dark and in the  hatched regions  $\phi_{ R k}$ with, respectively, positive and negative $k$ are constant. In the complements to those regions $\phi_{ L k}$ with $k$ of the
same sign are constant. \label{fig:karta}}
\end{figure}
\begin{teor}{Example}\label{ex:mode}
Assume $Q=L$, $k=-3$. Then we, first, find that  $u^*_{ Q |k|}$ is the function  (we again perceive its domain as the segment $t=0$ with the values at the missing points $x=\pm\polu,\mu\polu $ being defined by continuity) equal to $\sqrt{12\pi}e^{3i\xi(x)}$ at  $x\in[-\polu,\mu\polu]$ and $-\sqrt{12\pi}$ otherwise.
Correspondingly, for every $p\in \mathcal M$  we define $\phi_{ L, -3}(p)$ to be equal to
\[ \Psi(0,u^*_{  L 3})= \left\{
                                   \begin{array}{ll}
                                     \sqrt{12\pi}e^{3i\xi(x_p)}, & \hbox{at } x_p\in[-\polu,\mu\polu]; \\
                                     -\sqrt{12\pi}, & \hbox{otherwise.}
                                   \end{array}
                                 \right.
\]
where $x_p$ is the $x$-coordinate of the point at which the null $\alpha$-directed geodesic  through $p$ meets the segment $t=0$. Whence, in particular, $\phi_{ L, -3}=-\sqrt{12\pi}$ in the entire non-hatcfed region in figure~\ref{fig:karta}. To write down the explicit expression for $\phi_{ L, -3}$ in the remaining part of $\mathcal M$ one replaces $x\dashrightarrow-\beta=x-t$ in the relevant $u$ and specifies the periodicity condition
\[
\phi_{ L, -3}=
\sqrt{12\pi}e^{\frac{3i \pi (\mu- 1 -2\beta/\polu)}{1+\mu}},\quad \beta \in
\left\{
  \begin{array}{ll}
    [-\mu\polu + \polu (1+\mu)n,\polu+ \polu (1+\mu)n], & \hbox{in } \mathcal{M}_L; \\
{}[-\polu + \polu n,\polu+ \polu n],\  \hbox{in } \mathcal{M}_T
& n\in \mathds{Z}.
  \end{array}
\right.
\]
\end{teor}

\begin{teor}{Remark}\label{rem:textb}
The problem obtained by restricting the consideration to the right (for definiteness) leg and choosing the vacuum to be that defined by the set of modes $\{\phi_{Rk}\}$ is well studied, see \cite{BirDav}.
The index $Q$ absent in that case 
  may seem to double ``the number" of modes (for $\mu=0$, say). Note, however, that the separation  between the frequencies  of the modes in the trousers is twice that between the ones in the cylinder.
\end{teor}

Now let the sought-for  Hilbert space $\Hi$ be defined as the completion of the linear span of the set \[\{\phi_{Rk}\}\cup \{\phi_{Lk}\}
,\quad  k =1,2\ldots
\] by the norm generated by the scalar product \eqref{eq:scalPr}. That it is indeed a scalar product (which requires positive definiteness) follows  from the easily verified equality
\begin{equation}\label{eq:u-phi}
\KG{\phi_{Qk} ,\phi_{\check Qn} }
=\delta_{kn}\delta_{\check QQ},\qquad
 \text{for all }\check Q,Q\quad k,n=1,2\ldots
 \end{equation}

According to the plan outlined in the end of section~\ref{sub:plan} it only remains to prove  \eqref{eq:poln}, which is done in the Appendix.

\subsection{ The  vacuum  SET}
The (non-renormalized) vacuum expectation value of the SET is
$
\Sc{T_{\mu\nu}}=\sum_n \hat T_{\mu\nu}[\phi_n],
$
where\begin{equation}\label{eq:Tmode}
\hat T_{\mu\nu}[f]\obozn {f},_\mu {f^*},_\nu  -\tfrac12\eta_{\mu\nu}\eta^{\kappa \lambda } {f},_\kappa
{f^*},_\lambda
\end{equation}
 and the summation is over all modes \cite{BirDav}. The series diverges of course and to renormalize the   result one  introduces a cut-off factor  into the divergent sum by replacing $t\dashrightarrow t - i\delta$  and lets $\delta \to 0$ at the end of the calculation, that is, after subtracting the limit at $\polu\to \infty$. Thus,
\begin{equation}\label{eq:Tsum}
< \mathbf{T}^\text{ren}_{\mu\nu }>=
{T^L_{>}}_{\mu\nu} +{ T^L_{<}}_{\mu\nu} +  {T^R_{>}}_{\mu\nu} + { T^R_{<}}_{\mu\nu}
\end{equation}
where
\begin{equation*}
{T^Q_{\lessgtr}}_{\mu\nu}\obozn \lim_{\delta\to 0}\Bigl[  \Bigl(1- \lim_{\polu\to \infty}\Bigr)
\sum_{n\lessgtr 0} {\hat T}{}_{\mu\nu}[ \phi_{Qn},\delta]\Bigr],
\end{equation*}
where ${\hat T}{}_{\mu\nu}[ \phi_{Qn},\delta]$ is the result of the substitution $t\dashrightarrow t - i\delta$ into ${\hat T}{}_{\mu\nu}[ \phi_{Qn}]$.
Let us find the four terms in turn. In the coordinate basis
\begin{equation*}
    {\hat T}{}_{\mu\nu}\bigl[e^{-iC(t-x)}\bigr]= C^2
\begin{pmatrix}
   1&-1\\
-1&1
\end{pmatrix}
\end{equation*}
where $C$ is a real constant [note that the second term of \eqref{eq:Tmode} vanishes]. Correspondingly, the term ${T^L_{<}}_{\mu\nu}$, when non-zero (i.~e. in the hatched region in figure~\ref{fig:space_1}
) is
\begin{multline*}
    \lim_{\delta\to 0}\left\{ \bigl(1- \lim_{\polu\to \infty}\bigr)
\sum_{n< 0}{\hat T}{}_{\mu\nu}\left[\frac{  \exp\bigl\{i|n| \frac{\pi}{1+\mu} (\mu- 1 -\tfrac{2}{\polu}(\beta-i\delta) )\bigr\}  }{\sqrt{4\pi|n|}}      \right]\right\}
\\=
-\frac{ \pi}{12\polu^2(1+\mu)^2}
\begin{pmatrix}
   1&-1\\
-1&1
\end{pmatrix}.
 \end{multline*}
The term ${T^L_{>}}_{\mu\nu}$ differs from ${T^L_{<}}_{\mu\nu}$ in two respects: 1) it vanishes in the light gray region and 2) the exponents in the expression for modes depend now on $t+x$ instead of $t-x$, which changes the sign of the $tx$ components. Correspondingly,
\begin{equation*}
{T^L_{>}}_{\mu\nu}= -
\frac{ \pi}{12\polu^2(1+\mu)^2}
\begin{pmatrix}
   1&1\\
1&1
\end{pmatrix}\qquad \text{in the dark  region and $=0$ otherwise.}
\end{equation*}
Changing the sign of $\mu$
we find
\begin{multline*}\label{eq:TR<}
    {T^R_{<}}_{\mu\nu}=
-\frac{ \pi}{12\polu^2(1-\mu)^2}
\begin{pmatrix}
   1&-1\\
-1&1
\end{pmatrix}\qquad \text{in the non-hatched  region and $=0$ otherwise.}
\end{multline*}
\begin{equation*}\label{eq:TR>}
   {T^R_{>}}_{\mu\nu}= -
\frac{ \pi}{12\polu^2(1-\mu)^2}
\begin{pmatrix}
   1&1\\
1&1
\end{pmatrix}\qquad \text{in the light  region and $=0$ otherwise.}
\end{equation*}
Summing these terms up one finally gets (the description and numbering refer to  figure~\ref{fig:karta}):
\begin{multline*}
\text{light non-hatched regions I}\qquad
\Sc{\mathbf{T}^\text{ren}_{tt }}=\Sc{\mathbf{T}^\text{ren}_{xx }}=-\frac{ \pi}{6\polu^2(1-\mu)^2},
\\
\Sc{\mathbf{T}^\text{ren}_{tx }} =\Sc{\mathbf{T}^\text{ren}_{xt }}= 0;
\end{multline*}
\begin{multline*}
\text{dark hatched  regions III}\qquad
\Sc{\mathbf{T}^\text{ren}_{tt }}=\Sc{\mathbf{T}^\text{ren}_{xx }}=-\frac{ \pi}{6\polu^2(1+\mu)^2},
\\
\Sc{\mathbf{T}^\text{ren}_{tx }} =\Sc{\mathbf{T}^\text{ren}_{xt }}=0;
\end{multline*}
\begin{multline*}
\text{light hatched  regions IV}\qquad
\Sc{\mathbf{T}^\text{ren}_{tt }}=\Sc{\mathbf{T}^\text{ren}_{xx }}=-\frac{ \pi}{6\polu^2}\frac{1+\mu^2}{ (1-\mu^2)^2} ,
\\
\Sc{\mathbf{T}^\text{ren}_{tx }} =\Sc{\mathbf{T}^\text{ren}_{xt }}
=-\frac{ \pi}{3\polu^2}\frac{\mu}{(1-\mu^2)^2};
\end{multline*}
\begin{multline*}
\text{and dark non-hatched  regions II}\qquad
\Sc{\mathbf{T}^\text{ren}_{tt }}=\Sc{\mathbf{T}^\text{ren}_{xx }}=-\frac{ \pi}{6\polu^2}\frac{1+\mu^2}{ (1-\mu^2)^2},
\\
\Sc{\mathbf{T}^\text{ren}_{tx }} =\Sc{\mathbf{T}^\text{ren}_{xt }}=\frac{ \pi}{3\polu^2}\frac{\mu}{(1-\mu^2)^2}.
\end{multline*}
Thus, in the reference frame of a free falling observer the stress-energy tensor is component-wise bounded even though in the general case  it suffers discontinuities of the first kind. If the legs are equal ($\mu=0$) there is no ``thunderbolt"---the SET is perfectly regular.

\section*{Acknowledgements}
This work was supported by RNP Grant
No.~15-02-06-818.

\appendix
\section{The relation between the classical and the quantum spaces}

By construction  \Cio\ is dense in \Hi\ (since all $\phi_{Qk}\in\Cio$). In this appendix a converse, in a sense, property \eqref{eq:poln} is established.

For an arbitrary  function  $A\in   \Fi$ let us prove that
\begin{equation}\label{eq:decomp}
A= A^+ +A^- , \qquad\text{where } A^+, (A^-)^*\in \Hi_F.
\end{equation}
Obviously, this  implies \eqref{eq:poln} for all $f$ of the type $f= \Psi(A, 0)$
[it suffices to set $f^\pm= \tilde\Psi(A^\pm, 0)$
where $ \tilde\Psi$  is the extension by continuity of $ \Psi$  to the entire
$\Hi_F\otimes \Hi_F$].
 The case of $f= \Psi(0, B)$ is perfectly analogous and $f$ of  the general type is just the sum of those two plus a constant.
So, the validity of \eqref{eq:decomp} will prove \eqref{eq:poln}.

\emph{Proof of \eqref{eq:decomp}.}
The mode  $u_1$ and  the function $A$ are smooth except    at $x=\mu\polu$ where either of them has a---non-zero in the case of $u_1$---jump in its first derivative see [\eqref{eqLperiodA}],
\[
w'(\mu\polu+0) - w'(\mu\polu-0)=w'(\polu) - w'(-\polu),
    \qquad \text{where\ } w\equiv A,u_1.
\]
 So, we can find a continuously differentiable linear combination
\begin{equation*}
    A_1 \in C^1, \quad A_1(\polu) =A_1(-\polu), \quad A'_1(\polu) =A'_1(-\polu),
    \qquad \text{where\ } A_1\equiv A- C_u u_1, \ C_u=const.
\end{equation*}
Evidently, \eqref{eq:decomp} is true iff  it is true with $A$ replaced by $A_1$. Thus it  involves no loss of generality to assume that $A \in C^1$.

Now, consider the Fourier coefficients
\[F_{L ,k}\obozn\frac{1}{2\pi}\int_{-\pi}^{\pi}\evalat{A}{x\in[-\polu,\mu\polu]}{}(x(\xi )) e^{ik\xi }\,\rmd \xi, \qquad
F_{R ,k}\obozn\frac{1}{2\pi}\int_{-\pi}^{\pi}\evalat{A}{x\in[\mu\polu,\polu]}{}(x(\zeta )) e^{ik\zeta }\,\rmd \zeta .
\]
 of (the restrictions of)  $A$. By \cite[n$^\circ$708]{Fiht3} it follows from the continuous differentiability \footnote{And equality $A'_1(\polu) =A'_1(-\polu)$. Actually, there are a few more requirements, but they are automatically satisfied by functions of \Ci.} of $A$
that
\begin{equation}\label{wq:Ubyv}
    F_{ Q,k }= O(k^{-3}).
\end{equation}
This rate of convergence implies that for some functions $X_{ Q }^{\pm}$
\begin{multline}\label{eq:limS}
S_{ Q }^{+}(k_0)\obozn\sum_{k=1}^{k_0} F_{ Q, k}u_{ Q k}\sqrt{4\pi  |k|}
\quad \text{ and }\quad
 S_{ Q }^-(k_0) \obozn\sum_{k=-k_0}^{ 1}F^*_{ Q ,k}u_{ Q,-k}\sqrt{4\pi  |k|}
\text{ converge} \\
 \text{uniformly to, respectively, } X_{ Q }^{+}  \text{ and } X_{ Q }^{-}
\end{multline}
and
\begin{equation}\label{eq:limS'}
\bigl[{S}_{ Q }^{\pm}(k_0) \bigr]'\text{ converges uniformly to }  Y_{ Q }^{\pm}\obozn \bigl[{{X^\pm}}_{ Q } \bigr]'.
\end{equation}

The expression $S_{L }^+ +{S^-}_{L }^* + \frac{1}{( 1 +\mu)\polu} \int_{-\polu}^{\mu\polu} A(x)\,\rmd x$ on the interval $x\in [-\polu, \mu\polu]$ is a partial sum of the Fourier series
of $\evalat{A}{x\in[-\polu,\mu\polu]}{}$. So, it converges to
the said function there while the first two terms tend, respectively, to $X_{L}^+ $ and ${X_{L}^-}^*$). On the other interval
 (i.~e. at $x\in [\mu\polu, \polu]$) it converges to $A(\mu\polu)$. Similar considerations apply to $S_{R}^+ +{S^-}_{R }^*$ and hence
 \begin{multline}\label{eq:S'}
\evalat{A}{x\in[-\polu,\polu]}{} = X_{L}^+ + {X_{L}^-}^* +  \frac{1}{( 1 +\mu)\polu} \int_{-\polu}^{\mu\polu} A(x)\,\rmd x + X_{R}^+ + {X_{R}^-}^*  + \frac{1}{( 1 -\mu)\polu}\int_{\mu\polu}^{\polu} A(x)\,\rmd x  - A(\mu\polu)
\\=(X_{L}^+ +X_{R}^+) + ({X_{L}^-} + {X_{R}^-})^*
\end{multline}
[the last equality follows from \eqref{eq:normF}].
 Comparing this with \eqref{eq:decomp} we see that the latter is proven once we show that
\begin{equation}\label{eq:incl}
   X_{ Q }^+,{X_{ Q }^-}^*  \in \Hi_F.
\end{equation}
So, recall that by construction $S_{ Q }^+,{S_{ Q }^-}^*\in\Hi_F$, see \eqref{eq:limS}. At the same time
by \eqref{eq:limS} and \eqref{eq:limS'}
\[\max |{X}_{ Q }^{\pm}- S_{ Q }^{\pm}(k_0)|,
\max |{X'}_{ Q }^{\pm}- {S'}_{ Q }^{\pm}(k_0)| \xrightarrow[k_0\to \infty]{} 0
\]
and hence $S_{ Q }^+$ and ${S_{ Q }^-}^*$ converge  to, respectively, $X_{ Q }^+$ and ${X_{ Q }^-}^*$ \emph{in the metric of} $\Hi_F$, see~\eqref{eq:scalProdSc}.
Thus, the containment  \eqref{eq:incl} follows from the completeness of the Hilbert space $\Hi_F$.

\end{document}